\documentclass[conference]{IEEEtran}
\IEEEoverridecommandlockouts
\usepackage{cite}
\usepackage{amsmath,amssymb,amsfonts}
\usepackage{algorithmic}
\usepackage{siunitx}
\usepackage{graphicx}
\usepackage{textcomp}
\usepackage{xcolor}
\usepackage[a4paper, total={184mm,239mm}]{geometry}
\def\BibTeX{{\rm B\kern-.05em{\sc i\kern-.025em b}\kern-.08em
        T\kern-.1667em\lower.7ex\hbox{E}\kern-.125emX}}

\usepackage{fancyhdr}
\usepackage[all]{background}
\usepackage{stackengine}
\setstackEOL{\\}
\setstackgap{L}{\normalbaselineskip}
\SetBgContents{\color{gray}{\tiny \Longstack{PREPRINT - accepted In Proceedings of the 2023 International Conference on Hardware/Software Codesign and System Synthesis}}}
\SetBgPosition{4.2cm,1cm}
\SetBgOpacity{1.0}
\SetBgAngle{0}
\SetBgScale{1.8}

\usepackage{pgfplots}
\usepackage{pgfplotstable}
\usetikzlibrary{patterns, arrows, pgfplots.groupplots,hobby}
\usetikzlibrary{patterns}
\usetikzlibrary{positioning}
\usetikzlibrary{pgfplots.groupplots}
\usetikzlibrary{plotmarks}
\usetikzlibrary{calc}
\usepgfplotslibrary{external}
\usepackage{balance}
\usepackage{caption}
\usepackage{xcolor}
\usepackage{colortbl}
\usepackage{rotating}
\usepackage{multirow}
\usepackage{booktabs}
\usepackage{bm}
\usepackage{subfig}
\usepackage{enumitem}
\usepackage{soul}
\usepackage[export]{adjustbox}
\usepackage{bigstrut}
\usepackage{graphicx}
\usepackage{amsmath}
\definecolor{cg1}{HTML}{E8F1FA}
\definecolor{cg2}{HTML}{C7DDF2}
\definecolor{cg3}{HTML}{8EBAE5}
\definecolor{cg4}{HTML}{407FB7}
\definecolor{cg5}{HTML}{00549F}
\definecolor{red}{HTML}{CC071E}
\usepackage{balance}
\usepackage{keyval}
\usepackage{epstopdf}
\usepackage{upgreek}
\usepackage{float}

\begin{document}
\bstctlcite{IEEEexample:BSTcontrol}
\title{Work-in-Progress: A Universal Instrumentation Platform for Non-Volatile Memories
\thanks{This work was funded by the Federal Ministry of Education and Research (BMBF, Germany) in the project NEUROTEC II (16ME0398K, 16ME0399).}
\vspace{-4mm}
}
\author{
   \IEEEauthorblockN{Felix Staudigl\IEEEauthorrefmark{1},
                     Mohammed Hossein\IEEEauthorrefmark{1},
                     Tobias Ziegler\IEEEauthorrefmark{3},
                     Hazem Al Indari\IEEEauthorrefmark{1},
                     Rebecca Pelke\IEEEauthorrefmark{1},\\
                     Sebastian Siegel\IEEEauthorrefmark{4},
                     Dirk J. Wouters\IEEEauthorrefmark{4},
                     Dominik Sisejkovic\IEEEauthorrefmark{2},
                     Jan Moritz Joseph\IEEEauthorrefmark{1}, and
                     Rainer Leupers\IEEEauthorrefmark{1}
                     }
\vspace{+2mm}
   \IEEEauthorblockA{\IEEEauthorrefmark{1}
   \textit{Institute for Communication Technologies and Embedded Systems, RWTH Aachen University, Germany}
   }
   \IEEEauthorblockA{\IEEEauthorrefmark{2}
   \textit{Security, Privacy, and Safety Research Group, Corporate Research Robert Bosch GmbH, Germany}
   }
   \IEEEauthorblockA{\IEEEauthorrefmark{3}
   \textit{Institut für Werkstoffe der Elektrotechnik 2 (IWE-2), RWTH Aachen University, Germany}
   }
   \IEEEauthorblockA{\IEEEauthorrefmark{4}
   \textit{Peter Grünberg Institute (PGI-10), Forschungszentrum Jülich GmbH, German}
   }
   \{staudigl, hossein, pelke, joseph, leupers\}@ice.rwth-aachen.de\\
   \{ziegler, wouters, siegel\}@iwe.rwth-aachen.de\\
   dominik.sisejkovic@de.bosch.com\\
\vspace{-9mm}
}

\maketitle

\begin{abstract}
    Emerging non-volatile memories (NVMs) represent a disruptive technology that allows a paradigm shift from the conventional von Neumann architecture towards more efficient computing-in-memory (CIM) architectures. Several instrumentation platforms have been proposed to interface NVMs allowing the characterization of single cells and crossbar structures. However, these platforms suffer from low flexibility and are not capable of performing CIM operations on NVMs. Therefore, we recently designed and built the NeuroBreakoutBoard, a highly versatile instrumentation platform capable of executing CIM on NVMs. We present our preliminary results demonstrating a relative error \(<5\%\) in the range of \SI{1}{\kilo\ohm} to \SI{1}{\mega\ohm} and showcase the switching behavior of a \(\text{HfO}_2\)/Ti-based memristive cell.
\end{abstract}

\begin{IEEEkeywords}
ReRAM, CIM, LIM, memristor, instrumentation\vspace{-3mm}
\end{IEEEkeywords}

\begin{figure}[b!]
    \centering
    \includegraphics[width=\columnwidth]{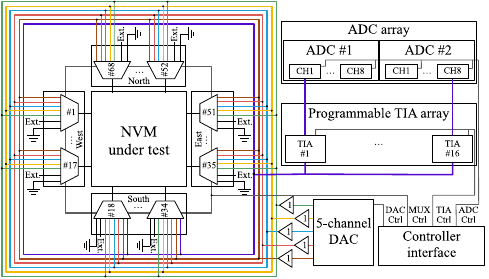}
    \caption{Overview of the implemented structure and modules.}
    \label{fig:neuro_arch}
    \vspace{-3mm}
\end{figure}

\section{Introduction}
Emerging non-volatile memories (NVMs) represent an ideal substrate for enabling computing-in-memory (CIM) by offering high density and non-volatility properties~\cite{Staudigl2022}. CIM can be implemented in analog~\cite{Ankit2019} or digital~\cite{Kvatinsky2014} fashion. Although the former offers the best computational efficiency, analog CIM requires expensive ADCs/DACs to convert inputs and outputs from the digital to the analog domain and vice versa. The latter uses so-called logic families to implement digital gates within the NVM, thereby circumventing the conversion but suffering from lower computational efficiency. Several instrumentation platforms have been proposed to characterize NVMs at the device and crossbar levels. However, most of these platforms only support passive crossbar structures~\cite{Berdan2015, Cayo2021, Fuente2020, Foster2020, Grossi2018, Xing2016, Xing2016a}. To the best of our knowledge, there are currently no platforms capable of executing neither analog nor digital CIM on NVMs. While the CIM operation is performed within the memory, the instrumentation platform must facilitate the surrounding circuitry to generate synchronized voltage pulses (inputs) and process the resulting currents (outputs). Hence, we designed and built the NeuroBreakoutBoard (NBB), a universal instrumentation platform to perform CIM on NVMs. Our platform supports all memristor-based memories in both passive (1R) and active (1T1R) crossbar configurations. Furthermore, the unique combination of the interconnection matrix, custom designed trans-impedance amplifiers (TIAs), and the implemented firmware enable the NBB to perform both analog and digital CIM operations by using different NVMs.

\begin{figure}[t!]
    \centering
    \includegraphics[width=\columnwidth]{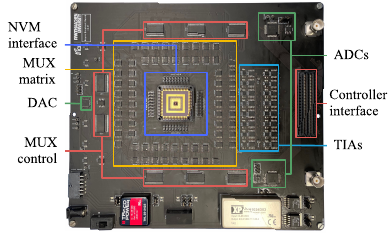}
    \caption{Image of the NeuroBreakoutBoard and its main components.}
    \label{fig:neuro_pic}
    \vspace{-7mm}
\end{figure}

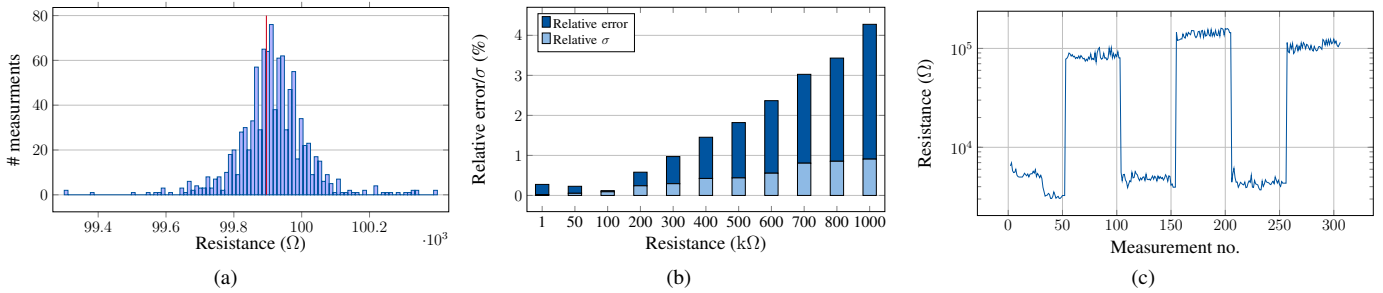
\begin{figure*}[t!]
    \centering
    \subfloat[]{
        \begin{tikzpicture}[scale=0.5]
        \begin{axis}[
            ybar,
            ylabel={\# measurments},
            xlabel={Resistance (\SI{}{\ohm})},
            ylabel style={at={(-0.01,0.5)},anchor=north},
            enlarge x limits={abs=0.3cm},
            ymajorgrids = true,
            scaled x ticks=base 10:-3,
            width=12cm, height=7cm,
            xtick={99400,99600,99800,100000,100200},
            xticklabel style={align=center},
            ylabel style ={font=\Large},
            xlabel style ={font=\Large},
            tick label style={font=\large},
            ]
            \addplot+[draw=cg5,hist={data min=99300, data max=100400, bins=100}] %
                table[x=Expected_Resistance,y=Measured_Resistance,col sep=comma] %
                {data/R_100_KOhm.csv};
            \addplot[red,thick,sharp plot,update limits=false,] %
                coordinates { (99896,0) (99896,80) };
        \end{axis}
        \end{tikzpicture}
    }
    \subfloat[]{
        \begin{tikzpicture}[scale=0.5]
        \begin{axis}[
            ybar stacked,
            width=11cm, height=7cm,
            enlarge x limits={abs=0.4cm},
            ylabel={Relative error/$\sigma$ (\%)},
            xlabel={Resistance (\SI{}{\kilo\ohm})},
            ymajorgrids = true,
            xtick=data,
            xticklabels={1,50,100,200,300,400,500,600,700,800,1000},
            legend style={cells={anchor=west}, legend pos=north west},
            reverse legend=true,
            xticklabel style={align=center},
            ylabel style ={font=\Large},
            xlabel style ={font=\Large},
            tick label style={font=\large},
        ]
        \addplot [fill=cg3] %
            table [y=r_std, x expr=\coordindex, col sep=comma] {data/error_analysis.csv};
        \addlegendentry{Relative $\sigma$}

        \addplot [fill=cg5] %
            table [y=r_error, x expr=\coordindex, col sep=comma] {data/error_analysis.csv};
        \addlegendentry{Relative error}

        \end{axis}
        \end{tikzpicture}
    }
    \hfil \subfloat[]{
        \begin{tikzpicture}[scale=0.5]
            \begin{semilogyaxis}[
                width=12cm,
                height=7cm,
                ylabel={Resistance (\SI{}{\ohm})},
                ylabel style={at={(-0.045,0.5)},anchor=north},
                ymajorgrids = true,
                ylabel style ={font=\Large},
                xlabel={Measurement no.},
                xmajorgrids = true,
                xlabel style ={font=\Large},
                tick label style={font=\large},
                ]
                \addplot[draw=cg5] %
                    table[x=measurement,y=resistance,col sep=comma] %
                    {data/mad200_switching.csv};
            \end{semilogyaxis}
        \end{tikzpicture}
    }
\caption{Result overview: (a) single device histogram (1,000 measurements), (b) relative error and standard deviation of various reference resistances, and (c) write/read operations performed on the MAD200 NVM chip.}\label{fig:results}
\vspace{-5mm}
\end{figure*}

\section{NeuroBreakoutBoard (NBB)}
The NBB generates the required input pulses with a multi-channel 12bit DAC (\SI{0}{\volt}-\SI{5}{\volt},\SI{0}{\volt}-\SI{10}{\volt}, \(\pm\)\SI{2.5}{\volt}, \(\pm\)\SI{5}{\volt}, \(\pm\)\SI{10}{\volt}). These inputs can be arbitrarily mapped by the interconnection matrix to the \textit{NVM interface} shown in Fig.~\ref{fig:neuro_pic}. The west/east/south multiplexers provide five independent potentials, one external measurement/supply line, and the ground potential. Additionally, the north multiplexers connect to the sensing module capable of measuring currents and voltages. The sensing module consists of an array of programmable TIAs, together with high-performance 14bit ADCs enabling precise measurements of resistances in the range of \SI{1}{\kilo\ohm} to \SI{1}{\mega\ohm}. Each TIA offers four distinct sensitivity levels subdividing the broad range of input currents/voltages by four to enhance the overall measurement accuracy. The NBB can control up to 68 lines simultaneously and connects via the \textit{NVM interface} to various extension boards implementing additional digital circuitry, chip packages, and measuring capabilities. The NBB is orchestrated by the \textit{controller interface}, which bundles all critical control and data signals. The \textit{controller unit} connects to this interface and provides a microcontroller which runs the implemented NBB firmware. The software offers a Python/C++ interface to issue write/read/compute operations to be executed on the crossbar array. The implemented firmware executes operations by assigning the respective voltage pulses on the corresponding pins. Likewise, the firmware controls the sensing module by iteratively adjusting the programmable TIAs and triggering the respective ADCs.

\vspace{-1mm}
\section{Result}
In this section, we discuss our preliminary results of two experiments that showcase the measurement accuracy and flexibility of the NeuroBreakoutBoard. Each of the four sensitivity stages has been calibrated with high-precision resistances to obtain maximum accuracy.

\textbf{Accuracy:} To determine the measurement accuracy of the NBB, we applied a similar methodology described by Berdan et al.~\cite{Berdan2015}. Fig.~\ref{fig:results} (a) illustrates the result of 1,000 measurements of a reference resistor. Each measurement incorporates 50 consecutive ADC samplings to omit the impact of noise. We determined the actual resistance of the resistor using the Keithley DMM7510 (marked red). Furthermore, Fig.~\ref{fig:results} (b) depicts the relative error and standard deviation of various reference resistances calculated based on 10,000 measurements per resistance. \textit{Both experiments report a high measurement precision (\(<5\%\) relative error and \(<1\%\) of relative \(\sigma\)) considering the additional line resistances and parasitic effects of the interconnection matrix.}

\textbf{Switiching characteristics:} Finally, we performed several write/read operations on a \(\text{HfO}_2\)/Ti 1T1R crossbar array. The crossbar is fabricated with the MAD200 process offered by CMP/CEA-LETI~\cite{Grossi2018a} in \SI{130}{\nano\meter} consisting of a demultiplexer and two \(512\times32\) crossbar structures. To facilitate the chip on the NBB, we manufactured an extension board connecting to the NVM interface. The extension board offers the required digital circuitry (I/O expanders) to drive and control the chip. Fig.~\ref{fig:results} (c) illustrates the resistance change of a single cell within the crossbar array over the course of alternating RESET/SET operations.

\section{Conclusion}
The NeuroBreakoutBoard is a versatile instrumentation platform to characterize NVMs and execute CIM operations. Our preliminary results indicate a relative error in the range of \SI{1}{\kilo\ohm} to \SI{1}{\mega\ohm} lower than \(5\%\) and a high precision (\(\sigma<1\%\)). Moreover, we performed several read/write operations on a 1T1R crossbar structure. In the future, we intend to provide the NeuroBreakoutBoard as a commercial solution, thereby facilitating the accessibility of memristor-based CIM to both academia and industry. Additionally, we aim to integrate the NeuroBreakoutBoard into a hardware-in-the-loop simulation environment to conduct thorough investigations on the reliability of non-volatile memories (NVMs) under realistic workloads.

\balance
\bibliographystyle{IEEEtran}
\bibliography{bibliography.bib}
\end{document}